\documentclass[12pt]{article}

\usepackage{graphicx}

\newcommand{\be}{\begin{equation}}
\newcommand{\ee}{\end{equation}}
\newcommand{\bea}{\begin{eqnarray}}
\newcommand{\eea}{\end{eqnarray}}
\newcommand{\beas}{\begin{eqnarray*}}
\newcommand{\eeas}{\end{eqnarray*}}

\newcommand{\DS}{{\Delta_{\textstyle *}}}

\begin{document}
\begin{titlepage}

\begin{center}

{\Large Does boundary quantum mechanics  \\ imply quantum mechanics in the bulk?}

\vspace{8mm}

\renewcommand\thefootnote{\mbox{$\fnsymbol{footnote}$}}
Daniel Kabat${}^{1}$\footnote{daniel.kabat@lehman.cuny.edu},
Gilad Lifschytz${}^{2}$\footnote{giladl@research.haifa.ac.il}

\vspace{4mm}

${}^1${\small \sl Department of Physics and Astronomy} \\
{\small \sl Lehman College, City University of New York, Bronx NY 10468, USA}

\vspace{2mm}

${}^2${\small \sl Department of Mathematics and} \\
{\small \sl Haifa Research Center for Theoretical Physics and Astrophysics} \\
{\small \sl University of Haifa, Haifa 31905, Israel}

\end{center}

\vspace{8mm}

\noindent
Perturbative bulk reconstruction in AdS/CFT starts by representing a free bulk field $\phi^{(0)}$ as a smeared operator in the CFT.  
A series of $1/N$ corrections must be added to $\phi^{(0)}$ to represent an interacting bulk field $\phi$.
These corrections have been determined in the literature from several points of view.
Here we develop a new perspective.
We show that correlation functions involving $\phi^{(0)}$ suffer from ambiguities due to analytic continuation.
As a result $\phi^{(0)}$ fails to be a well-defined linear operator in the CFT.
This means bulk reconstruction can be understood as a procedure for building up well-defined operators in the CFT which thereby singles out the interacting field $\phi$.
We further propose that the difficulty with defining $\phi^{(0)}$ as a linear operator can be re-interpreted as a breakdown of associativity.
Presumably $\phi^{(0)}$ can only be corrected to become an associative operator in perturbation theory.
This suggests that quantum mechanics in the bulk is only valid in perturbation theory around a semiclassical bulk geometry.

\end{titlepage}
\setcounter{footnote}{0}
\renewcommand\thefootnote{\mbox{\arabic{footnote}}}

\section{Introduction}
In AdS/CFT \cite{Maldacena:1997re} one identifies the Hilbert space of the CFT with the Hilbert space of the bulk theory.  To make the identification precise one needs to show
how one represents local bulk excitations in the CFT.  To this end an algorithm has been developed for representing local bulk fields
as operators in the CFT \cite{Kabat:2011rz}.\footnote{Some useful previous results on the construction of bulk observables are reviewed in appendix \ref{sect:review}.}
The algorithm works in $1/N$ perturbation theory and generates an expression for a local interacting bulk field as a sum
of CFT operators.
\be
\label{PhiSeries}
\phi = \phi^{(0)} + \phi^{(1)} + \phi^{(2)} + \cdots
\ee
Here $\phi^{(0)}$ is built by smearing a single-trace operator in the CFT over a region on the boundary.  For various approaches to constructing $\phi^{(0)}$ see
\cite{Banks:1998dd,Dobrev:1998md,Bena:1999jv,Hamilton:2005ju,Hamilton:2006az,Hamilton:2006fh,Verlinde:2015qfa,Miyaji:2015fia,Nakayama:2015mva,Kabat:2017mun,Faulkner:2017vdd,Cotler:2017erl}.
$\phi^{(1)}$ is an ${\cal O}(1/N)$ correction built by smearing double-trace operators, $\phi^{(2)}$ is an ${\cal O}(1/N^2)$ correction from triple-trace operators, and so on.
$\phi^{(0)}$ behaves like a free field in the bulk and can be used to reproduce bulk correlators to ${\cal O}(N^0)$ (bulk 2-point functions).  The sum $\phi^{(0)} + \phi^{(1)}$
can be used to reproduce bulk correlators to ${\cal O}(1/N)$ (bulk 3-point functions).  $\phi^{(0)} + \phi^{(1)} + \phi^{(2)}$ allows one to reproduce bulk 4-point functions to ${\cal O}(1/N^2)$, and so on.

Although the $1/N$ expansion of the CFT corresponds to the bulk expansion in powers of the gravitational coupling $\kappa \sim \sqrt{G_N}$, it's important to recognize that the
expansion (\ref{PhiSeries}) is not simply a re-writing of bulk perturbation theory.  In particular there is no division of the CFT into ``free'' and ``interacting'' parts.  Instead we always
work with a fixed interacting finite-$N$ CFT, and we define a series of operators $\phi^{(0)}$, $\,\phi^{(0)} + \phi^{(1)}$, $\,\phi^{(0)} + \phi^{(1)} + \phi^{(2)}$, $\ldots$ in this interacting
CFT whose correlators provide better and better approximations to bulk correlators.
For scalar fields the corrections to $\phi^{(0)}$ have been determined
by requiring that the CFT operator corresponding to the local bulk field obeys an appropriate notion of micro-causality\footnote{i.e.\ the property that field operators commute at
spacelike separation} when inserted in correlation functions \cite{Kabat:2011rz}. From the CFT perspective this
corresponds to certain analytic properties of these correlation functions.  Rather remarkably this condition is enough to determine the expression for local bulk operators in the CFT up to field redefinitions
and gives the correct bulk equations of motion at least at tree level \cite{Kabat:2015swa}.  For other approaches to interacting fields see \cite{Heemskerk:2012mn,Lewkowycz:2016ukf,Guica:2016pid,Anand:2017dav,Chen:2017dnl}.

This algorithm, while successful in reproducing bulk perturbation theory from the CFT, is still lacking in some respects.
First, the condition of bulk micro-causality, although it can be represented as an analytic property of correlators in the CFT, does not have a clear explanation in terms of CFT considerations per se.
Second, for scalars coupled to gauge fields (or gravity), the correlator of a bulk scalar with a boundary current (or the stress tensor) does not obey a simple notion of bulk micro-causality.
This is due to Gauss constraints which demand Wilson lines stretching out to the boundary.  While this obstacle can be overcome by identifying components of the boundary field strengths for which
micro-causality holds \cite{Kabat:2012av,Kabat:2013wga}, this makes it clear that bulk micro-causality is a useful but not fundamental criterion.  Moreover, if one considers bulk gauge fields and gravitons, then even at the level of two-point functions correlators do not obey bulk micro-causality \cite{Heemskerk:2012np,Kabat:2012hp,Sarkar:2014dma}.  So correcting CFT operators in order to preserve micro-causality in the presence of interactions
is a somewhat dubious procedure.

In fact one could ask a broader question.  Why should the bulk dual of the CFT be a local theory at all?  After all complete information about the bulk is captured by the CFT whose correlation functions carry all the required data. The only criterion that seems necessary is that bulk correlation functions should approach CFT correlation functions (up to factors $\sim Z^\Delta$) as the bulk points approach the boundary.  One can imagine many bulk correlation functions consistent with this condition.  For instance one could just use $\phi^{(0)}$ to produce ``bulk correlation functions.''
Of course the resulting bulk theory won't be local, but from a CFT perspective what's wrong with this procedure?

To address this in section \ref{sect:ambiguities} we show that correlation functions involving $\phi^{(0)}$ suffer from ambiguities due to analytic continuation.  We then present some
implications of this result.
\begin{itemize}
\item
In \S \ref{sect:obstruction} we show that in general $\phi^{(0)}$ cannot be realized as a linear operator on a Hilbert space.
\item
In \S \ref{sect:reconstruction} we discuss implications of this for the program of perturbative bulk reconstruction, which can now be understood as a procedure for building up
well-defined CFT operators.  We show that this new paradigm for reconstruction has practical advantages, in particular that it applies without modification to bulk fields with gauge redundancy.
\item
In \S \ref{sect:associativity} we draw broader conclusions about bulk physics.  Given the difficulty with defining $\phi^{(0)}$ one possible response would be to abandon any bulk
interpretation of $\phi^{(0)}$, and to assert that only the corrected field $\phi$ has meaning in the bulk.\footnote{In the language of
quantum error correction \cite{Almheiri:2014lwa,Mintun:2015qda}, only $\phi$ would be a logical operator defined on the code subspace.}  As an alternative response, we advocate that
correlators involving $\phi^{(0)}$ can be given a consistent bulk interpretation, in which however the fundamental property of operator associativity is violated.  In other words, we are proposing that
$\phi^{(0)}$ can be understood as a logical but non-associative operator.  Since $\phi^{(0)}$ presumably can only be corrected to become associative in perturbation theory, this suggests that bulk quantum
mechanics only emerges in perturbation theory around a given semiclassical bulk geometry.
\end{itemize}

\section{Ambiguities in correlators with $\phi^{(0)}$\label{sect:ambiguities}}
In this section we study 3-point correlators involving $\phi^{(0)}$  and two boundary operators and show that ambiguities arise due to analytic continuation.  We will treat the simple case of massless fields in AdS${}_3$.  The results generalize to arbitrary spacetime and conformal dimensions, as shown in appendix \ref{sect:GeneralDimensions}.

Consider a massless scalar field $\phi$ in AdS${}_3$ which is dual to a primary scalar operator ${\cal O}$ of dimension
$\Delta = 2$ in the CFT.  The CFT 3-point function is, with coefficient $\gamma$
\be
\label{CFT3point}
\langle {\cal O}_1 {\cal O}_2 {\cal O}_3 \rangle = \gamma \prod_{i<j} {1 \over -(T_i - T_j - i \epsilon_{ij})^2 + (X_i - X_j)^2}
\ee
where we are abbreviating
\be
{\cal O}_1 \equiv {\cal O}(T_1,X_1) \qquad\,\, {\cal O}_2 \equiv {\cal O}(T_2,X_2) \qquad\,\, {\cal O}_3 \equiv {\cal O}(T_3,X_3)
\ee
This is a Lorentzian Wightman correlator, with the operators in the indicated order, corresponding to an $i \epsilon$ prescription
with $\epsilon_1 > \epsilon_2 > \epsilon_3$ approaching $0^+$ and $\epsilon_{ij} = \epsilon_i - \epsilon_j$.  We can smear one of the operators into the bulk by defining
\be
\label{MasslessAdS3Smear}
\phi^{(0)}(T,X,Z) = {1 \over 2\pi} \int\limits_{(T')^2 + (Y')^2 < Z^2} \hspace{-5mm} dT' dY' \, {\cal O}(T + T',X + i Y')
\ee
Here $(T,X,Z)$ labels a point in the Poincar\'e patch of AdS${}_3$ with metric
\[
ds^2 = {R^2 \over Z^2}\left(-dT^2 + dX^2 + dZ^2\right)
\]

Following \cite{Kabat:2011rz} we can compute the mixed bulk -- boundary 3-point function $\langle \phi^{(0)} {\cal O}_2 {\cal O}_3 \rangle$ by applying the smearing integral (\ref{MasslessAdS3Smear})
to the CFT correlator (\ref{CFT3point}).  Since the CFT correlator has singularities it is not a priori obvious that we will obtain a well-defined result.
A reasonable prescription is to start near the boundary $Z \rightarrow 0$, since in this limit the smearing region shrinks to a point
and we must recover the correlator of local operators in the CFT.  Then we can analytically continue to finite $Z$, taking care with any singularities we encounter along the way.

This procedure was followed in \cite{Kabat:2011rz}, with the result
\be
\label{PhiOO}
\langle \phi^{(0)} {\cal O}_2 {\cal O}_3 \rangle = {\gamma \over 2 \left(T_{23}^+ T_{23}^-\right)^2} \log {\chi \over \chi - 1}
\ee
where we're abbreviating
\be
\phi^{(0)} \equiv \phi^{(0)}(T_1,X_1,Z_1)
\ee
We're using light-front coordinates on the boundary, $T^{\pm} = T \pm X$ and $T_{ij} = T_i - T_j$, and we've introduced an AdS-invariant quantity $\chi$ defined by
\be
\label{ChiOverChi-1}
{\chi \over \chi - 1} = {\left(-T_{12}^+T_{12}^- + Z_1^2\right) \left(-T_{13}^+T_{13}^- + Z_1^2\right) \over \left(-T_{12}^+T_{13}^- + Z_1^2\right) \left(-T_{13}^+T_{12}^- + Z_1^2\right)}
\ee
This quantity inherits an $i \epsilon$ prescription from the boundary Wightman correlator, namely $T_i \rightarrow T_i - i \epsilon_i$.  With operators ordered
as in (\ref{PhiOO}) we would have $\epsilon_1 > \epsilon_2 > \epsilon_3$.  Of course other orderings are possible.  For example $\langle {\cal O}_2 \phi^{(0)} {\cal O}_3 \rangle$
corresponds to $\epsilon_2 > \epsilon_1 > \epsilon_3$.

The correlator (\ref{PhiOO}) is singular at $\chi = 0$ and $\chi = 1$, with a branch cut along the real axis for $0 < \chi < 1$.  Our goal is to determine
the implications of these singularities.  There are two cases to consider: boundary operators that are spacelike separated and boundary operators
that are timelike separated.  Here we will treat spacelike separated operators.  For timelike separation see appendix \ref{sect:timelike}.

\subsection{Spacelike separated boundary points\label{sect:spacelike}}
\begin{figure}
\center{\includegraphics{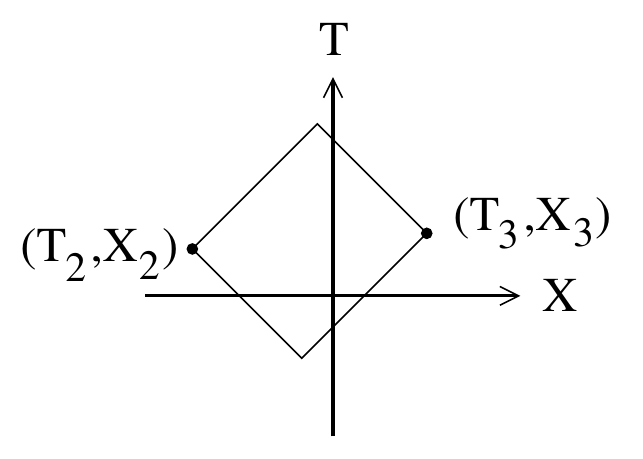} \hfil \includegraphics{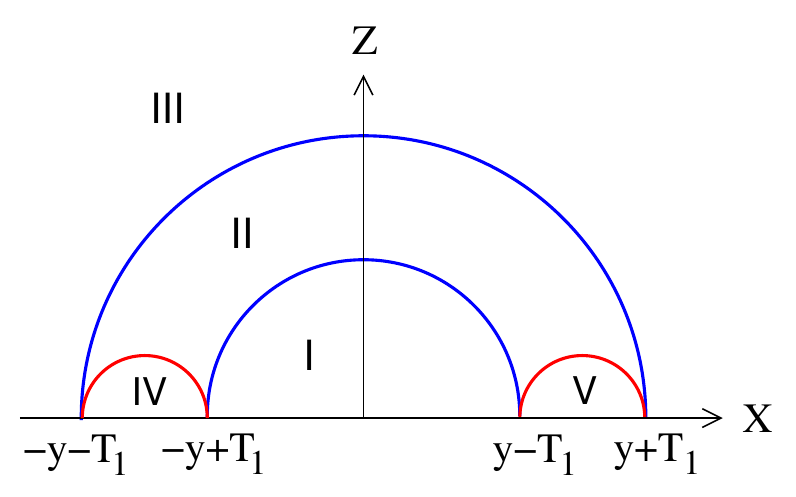}}
\caption{On the left, the causal diamond on the boundary determined by two spacelike separated points.  On the right, a bulk time slice with various regions indicated.
The two red curves correspond to $\chi = 0$ and the two blue curves correspond to $\chi = 1$.  In region {\sf II} we have $0 < \chi < 1$.\label{fig:spacelike}}
\end{figure}

Suppose the two boundary points $(T_2,X_2)$ and $(T_3,X_3)$ are spacelike separated.  These points define a causal diamond
on the boundary as shown in Fig.\ \ref{fig:spacelike}.
\begin{itemize}
\item
For points in the bulk with $\chi = 0$ one of the factors in the numerator of (\ref{ChiOverChi-1}) vanishes, so such points are light-like separated from either the left or right corner
of the diamond.  Such singularities are expected, since the bulk point is null separated from one of the boundary operators.
\item
Points in the bulk with $\chi = 1$ are light-like separated from either the top or bottom corner of the diamond.  To see this note for example that the top of the diamond
has light-front coordinates $T^+ = T_3^+$, $T^- = T_2^-$ and null separation from this point corresponds to $-T_{13}^+ T_{12}^- + Z_1^2 = 0$.  There are no operators at the top or bottom of the
diamond, so such singularities are {\em not} anticipated based on the bulk causal structure.
\end{itemize}

Note that the two light-cones which make up the $\chi = 1$ surface intersect on a semicircle in the bulk.  This semicircle has a geometric interpretation as a spacelike geodesic connecting
the two boundary operators.  Thus it can be thought of as the RT surface appropriate for the boundary interval between the two operators \cite{Ryu:2006bv}.  It plays a further role since
the OPE between ${\cal O}_2$ and ${\cal O}_3$ when projected on a conformal family can be written as the integral of $\phi^{(0)}$ along the geodesic \cite{Hijano:2015zsa,Czech:2016xec, daCunha:2016crm, deBoer:2016pqk}.\footnote{The RT property is special to AdS${}_3$ but the OPE property holds in any dimension.}  This makes it clear why the correlator (\ref{PhiOO}) is singular at $\chi = 1$:
as $\chi \rightarrow 1^+$ one encounters a singularity as the bulk $\phi^{(0)}$ becomes null separated from a point on the geodesic.  Note that points with $0 < \chi < 1$ are to the future or past of the geodesic.

Let's consider the analytic structure of the correlator (\ref{PhiOO}) in more detail.  As mentioned above, (\ref{PhiOO}) was obtained by performing the smearing integral near $Z = 0$
and analytically continuing the result into the bulk.  However we should ask: does the answer only depend on the final bulk point?  Or does it depend on the path that was
used to get there?

Since we are considering boundary points at spacelike separation, without loss of generality we place the boundary operators at $(T_2 = 0,\,X_2 = -y)$ and $(T_3 = 0,\,X_3 = +y)$.
The bulk point is $(T_1,X_1,Z_1)$.  For simplicity we will restrict our analysis to bulk points with $0 < T_1 < y$.  The correlator (\ref{PhiOO}) is then\footnote{Operator ordering in this expression is determined by the values
of $\epsilon_i$.  Operators are ordered so that ${\rm Im} \, T$ increases from left to right.}
\be
\langle \phi^{(0)} {\cal O}_2 {\cal O}_3 \rangle = {\gamma \over 2 \left(T_{23}^+ T_{23}^-\right)^2} \log {AB \over CD}
\ee
where
\beas
\label{ABCD}
A & = & -T_1^2 + (X_1 + y)^2 + Z_1^2 + 2 i \epsilon_{12} T_1 \\
B & = & -T_1^2 + (X_1 - y)^2 + Z_1^2 + 2 i \epsilon_{13} T_1 \\
C & = & -(T_1+y)^2+X_1^2+Z_1^2+i \epsilon_{12}(T_1+y-X_1) + i \epsilon_{13}(T_1+y+X_1) \\
D & = & -(T_1-y)^2+X_1^2+Z_1^2+i \epsilon_{12}(T_1-y+X_1) + i \epsilon_{13}(T_1-y-X_1)
\eeas

To see the significance of these formulas a timeslice containing the bulk point is shown in Fig.\ \ref{fig:spacelike}.
The curves $A = 0$ and $B = 0$ are shown in red; they indicate where $\chi = 0$.  The curves
$C = 0$ and $D = 0$ are shown in blue; they indicate where $\chi = 1$.  In the regions near the boundary the argument of the log is positive, but in region {\sf II} which corresponds
to $0 < \chi < 1$ it is negative.
Crucially in region {\sf II} the argument of the log has an imaginary part ${\rm Im} \, {\rm Arg}$ which depends on the analytic continuation.  That is
it depends on the starting point, which determines which of the factors $A$, $B$, $C$, $D$ crosses zero as we
continue into region {\sf II}.  Up to a product of positive quantities which don't vanish at the singularity, the outcomes are shown in Table \ref{SpacelikeEpsilons}.

\begin{table}
\begin{center}
\begin{tabular}{c|c|l}
start from region & cross & \quad resulting $i \, {\rm Im} \, {\rm Arg}$ \\
\hline
\vphantom{\Big(}{\sf IV} & $\chi = 0$ at $A = 0$ & $-2 i \epsilon_{12} T_1$ \\
{\sf V} & $\chi = 0$ at $B = 0$ & $-2 i \epsilon_{13} T_1$ \\
{\sf I} & $\chi = 1$ at $D = 0$ & $+i \epsilon_{12}(T_1-y+X_1) + i \epsilon_{13}(T_1-y-X_1)$ \\
{\sf III} & $\chi = 1$ at $C = 0$ & $-i \epsilon_{12}(T_1+y-X_1) - i \epsilon_{13}(T_1+y+X_1)$
\end{tabular}
\end{center}
\caption{Continuing into region {\sf II} for the case of spacelike separated boundary operators.\label{SpacelikeEpsilons}}
\end{table}

The factors multiplying the $\epsilon_{ij}$'s in Table \ref{SpacelikeEpsilons} are to be evaluated where the singularity is crossed.  Fortunately
these factors all have definite signs at these points.  So up to rescalings by position-dependent but positive quantities we can summarize the
outcome as
\begin{center}
\begin{tabular}{c|c}
start from region & $i \, {\rm Im} \, {\rm Arg}$ \\
\hline
\vphantom{\Big(}{\sf IV} & $-i \epsilon_{12}$ \\
{\sf V} & $-i \epsilon_{13}$ \\
{\sf I} & $-i \epsilon_{12} - i \epsilon_{13}$ \\
{\sf III} & $-i \epsilon_{12} - i \epsilon_{13}$
\end{tabular}
\end{center}

At this stage we have cause for concern since the bulk correlator we have obtained depends not only on the final bulk point but also on the path chosen for analytic continuation.
The consequences of this depend on how the operators are ordered.

\noindent
\underline{\em bulk operator on left} \\
\noindent
To place the bulk operator on the left we take $\epsilon_{12} > 0$ and $\epsilon_{13} > 0$.  In this case the starting region doesn't matter.  No matter what path we follow, we
end up in region {\sf II} with ${\rm Im} \, {\rm Arg} = 0^-$.

\noindent
\underline{\em bulk operator on right} \\
\noindent
To place the bulk operator on the right we take $\epsilon_{12} < 0$ and $\epsilon_{13} < 0$.  Again the starting region doesn't matter, and no matter what path we follow we
end up in region {\sf II} with ${\rm Im} \, {\rm Arg} = 0^+$.

\noindent
\underline{\em bulk operator in the middle} \\
\noindent
For the ordering $\langle 0 \vert {\cal O}_2 \phi^{(0)} {\cal O}_3 \vert 0 \rangle$ we take $\epsilon_{12} < 0$ and $\epsilon_{13} > 0$.  Now the starting region makes a
difference.  If we start from region {\sf IV}, that is timelike to ${\cal O}_2$, then ${\rm Im} \, {\rm Arg} = 0^+$.  On the other hand if we start from region {\sf V}, that is
timelike to ${\cal O}_3$, then ${\rm Im} \, {\rm Arg} = 0^-$.  The continuations from regions {\sf I} and {\sf III} are ill-defined in the sense that ${\rm Im} \, {\rm Arg}$ depends
on the relative size of $\epsilon_{12}$ and $\epsilon_{13}$ and also on the values of $T_1,X_1,y$ at the point where the singularity is crossed.  This makes continuation from
these regions ambiguous.

\noindent
There are a few comments on correlators in region {\sf II} that are worth making at this point.  First, note that if the bulk operator is on the left or right the correlator is
unambiguous but comes with opposite $i \epsilon$ prescriptions.  Since the commutator is then given by the discontinuity across the cut of the log, we can say with certainty that in region {\sf II}
\be
\langle \, [ \phi^{(0)}, {\cal O}_2 {\cal O}_3 ] \, \rangle \not= 0
\ee
If the bulk operator is in the middle the situation is more subtle.
Note that starting from region {\sf IV}, i.e.\ timelike to ${\cal O}_2$, gives the same $i \epsilon$ prescription as if the bulk operator was on the right, so for this particular
continuation we would have
\be
\label{IVcommutator}
\langle {\cal O}_2 \phi^{(0)} {\cal O}_3 \rangle_{\sf IV} = \langle {\cal O}_2 {\cal O}_3 \phi^{(0)} \rangle
\ee
On the other hand starting from region {\sf V}, i.e.\ timelike to ${\cal O}_3$, gives the same prescription as placing the bulk operator on the left, so for this particular
continuation we would have
\be
\label{Vcommutator}
\langle {\cal O}_2 \phi^{(0)} {\cal O}_3 \rangle_{\sf V} = \langle \phi^{(0)} {\cal O}_2 {\cal O}_3 \rangle
\ee
We will return to interpretation of these results below, around equations (\ref{NonAssoc[]1}) and (\ref{NonAssoc[]2}).

\bigskip
\centerline{* \hspace{3cm} * \hspace{3cm} *}

To summarize, we've identified a region of the bulk where correlators must be defined by analytic continuation.
If the bulk operator is on the left or right the continuation is unambiguous in the sense that the final correlator one
obtains is independent of the path that is followed.  But if the bulk operator is in the middle the continuation depends
on the path.  Crossing the light-cone of a boundary operator ($\chi = 0$) is well-defined but the outcome depends on which light-cone
is crossed.  Crossing a $\chi = 1$ singularity is ambiguous and does not lead to a well-defined correlator.

In appendix \ref{sect:timelike} we show that the statements in the previous paragraph are also valid when the boundary points are timelike separated in AdS${}_3$.  In
appendix \ref{sect:GeneralDimensions} we show that they continue to hold for general spacetime and conformal dimensions.  Correlators generically have
a branch cut for $0 < \chi < 1$, and continuation into this region proceeds in a manner very similar to what we found for massless fields in AdS${}_3$.

\section{Implications\label{sect:implications}}
We conclude by discussing the implications of our results.  First we display a precise sense in which $\phi^{(0)}$ cannot be regarded as a well-defined CFT operator.
Then we present a new paradigm in which bulk reconstruction becomes a perturbative algorithm for generating well-defined observables in the CFT.  Finally we
propose that $\phi^{(0)}$ can be given an operator interpretation, in which however the fundamental property of operator associativity is lost.

\subsection{Obstruction to an operator interpretation of $\phi^{(0)}$\label{sect:obstruction}}
To summarize the previous section, we've identified a region of the bulk $0 < \chi < 1$ where the correlator of $\phi^{(0)}$ with two boundary operators
must be defined by analytic continuation.  One can enter this region by crossing $\chi = 1$.  However the singularities at $\chi = 1$ are particularly problematic
since they make the analytic continuation ambiguous.  They introduce dependence on the relative sizes of the $\epsilon_{ij}$, on the
position in the bulk where the $\chi= 1$ singularity is crossed, and on the positions of other operators in the correlator.  This can be seen explicitly
in Table \ref{SpacelikeEpsilons} and \ref{TimelikeEpsilons}.  In the presence of a singularity at $\chi = 1$ the singularities at $\chi = 0$ can
also cause problems.\footnote{See the discussion at the start of the next section.}  They do not obstruct analytic continuation, since they correspond to crossing the light-cone of a boundary operator, but they can assign inconsistent $i \epsilon$ prescriptions
depending on which light-cone is crossed.

What does this imply for the idea that $\phi^{(0)}$ could be an operator in the CFT?  To focus the discussion note that $\phi^{(0)}$ can be formally represented either as a smeared
CFT operator or as an infinite sum of local CFT operators.  For example in AdS${}_3$ \cite{daCunha:2016crm}
\be
\phi^{(0)} = \int K {\cal O} = {Z^\Delta \over 2} \sum_{m=0}^{\infty}\frac{\Gamma(\Delta-1)Z^{2m}}{\Gamma(m+1)\Gamma(m + \Delta)}(\partial_+\partial_-)^{m}{\cal O}(T^+,T^-)
\ee
When used inside a correlator, as long as the sum is convergent there is no problem.\footnote{Equivalently, as long as the integral encounters no singularities there is no problem.}
But of course the sum may not converge.  Then analytic continuation becomes necessary and can generate expressions that depend on additional data.  We performed such a continuation
in the previous section and found, in particular, that {\em when the bulk operator is in the middle of the correlator the continuation depends on the other operators in the correlator.}  This can be seen in Table
\ref{SpacelikeEpsilons} (or \ref{TimelikeEpsilons}), where crossing $\chi = 1$ with the bulk operator in the middle gives an overall $i \epsilon$ prescription that depends explicitly on $y$ (or $t$).
Even though it's not manifest in the formulas, crossing $\chi = 0$ with the bulk operator in the middle also depends on the other operators in the correlator.  This is due to the fact that
to cross $\chi = 0$ one must start on the boundary at timelike separation from one of the other operators, a condition which obviously depends on the positions of the other operators.

We see that a prescription for calculating correlators involving $\phi^{(0)}$ through analytic continuation from the boundary cannot be made well-defined in advance, without knowing what other
operators will be inserted in the correlator.  But we will insist that a good CFT operator can be characterized intrinsically and has correlators which can be uniquely defined without such advance
knowledge.  By this criterion we conclude that $\phi^{(0)}$ is not well-defined as a CFT operator.

\subsection{Implications for bulk reconstruction\label{sect:reconstruction}}
We've argued that $\phi^{(0)}$ is not well-defined as a CFT operator.  We now show that this gives a new perspective on bulk reconstruction, as an algorithm for correcting free
bulk operators such as $\phi^{(0)}$ in perturbation theory so that they have well-defined correlators.  This requirement applies to all bulk fields, including fields with gauge redundancy,
and appears to be a universal approach to bulk reconstruction.

First note that the difficulty with defining $\phi^{(0)}$ is tied to the nature of the $\chi = 1$ singularity.  A crucial property which allows the analytic continuation to be ambiguous
is that together the surfaces $\lbrace \chi = 0 \rbrace \cup \lbrace \chi = 1 \rbrace$ wall off a region of the bulk from the boundary.  This enables continuation from the boundary
into the region $0 < \chi < 1$ to be ambiguous.  Thus the algorithm for correcting a bulk operator so that it becomes well-defined must proceed by eliminating the branch point at
$\chi = 1$.  Indeed for cubic scalar couplings a set of corrected operators were worked out in \cite{Kabat:2011rz, Kabat:2015swa} with the result that the 3-point function of
a corrected operator with two boundary operators is analytic for $\chi>0$.  This was done with the idea that the $\chi=1$ singularity is an obstruction to bulk locality.  Here we are
re-interpreting it as an obstruction to $\phi^{(0)}$ being a well-defined CFT operator.

For a bulk scalar interacting with a gauge field or gravity the construction of corrected bulk operators was carried out in \cite{Kabat:2013wga, Kabat:2015swa}.
The lowest order operator $\phi^{(0)}$ suffers from the same problem as above inside a 3-point function with a boundary conserved current (or the energy momentum tensor) and a primary scalar.
A corrected bulk operator was constructed such that its 3-point function with $F_{\mu \nu}$ and another scalar operator, or with $C_{\alpha \beta \gamma \delta}$ and another scalar operator,
is analytic for $\chi>0$.  Here $F_{\mu\nu}$ and $C_{\alpha\beta\gamma\delta}$ are the boundary field strength and Weyl tensor.\footnote{$F_{\mu \nu}=\partial_{\mu}j_{\nu}-\partial_{\nu}j_{\mu}$ and  $C_{\alpha \beta \gamma \delta}= \partial_{\alpha}\partial_{\gamma}T_{\beta \delta}-\partial_{\alpha}\partial_{\delta}T_{\beta \gamma}-\partial_{\beta}\partial_{\gamma}T_{\alpha \delta}+\partial_{\beta}\partial_{\delta}T_{\alpha \gamma}$.}
The corrected bulk scalar operator solves the expected bulk equation of motion in holographic gauge \cite{Kabat:2015swa}, so its 3-point function will give the expected result.
But note that if one looks at the the 3-point function of the corrected bulk scalar with a boundary scalar and either $j_{\nu}$ or $T_{\mu \nu}$ then the result
will be non-local.\footnote{So there is still non-analyticity at spacelike separation, but it is not in the problematic form of a singularity at $\chi = 1$.}  This must be so due to Gauss law constraints.  So correlators of the corrected field, although they are well-defined, are not local.  This is an example where
corrections are necessary, not so much to restore locality, as to make correlators well-defined.
  
For bulk gauge fields and metric perturbations only some of the computations have been done.  Bulk 2-point functions already do not obey bulk micro-causality.  Inside 3-point functions the
lowest-order operator $F_{\mu\nu}^{(0)}$ suffers from the same disease of a branch cut for $0<\chi<1$.  We conjecture that the addition of higher-dimension
multi-trace operators in this case is necessary so that bulk 3-point functions are well-defined.  Once the 3-point function is well-defined, the extension to higher-point functions seems
to be possible using a bootstrap approach, at least at tree level \cite{Kabat:2016zzr}.

\subsection{Associativity and bulk quantum mechanics\label{sect:associativity}}
We've argued that $\phi^{(0)}$ is not a well-defined CFT operator.  An immediate reaction might be to give up on $\phi^{(0)}$ as a meaningful object, and to say that only the corrected
bulk operators have a sensible interpretation.  Here we will propose that $\phi^{(0)}$ can be given a consistent interpretation, in which however the fundamental property of operator
associativity is abandoned.

Let's look at the evidence we have.
\begin{itemize}
\item
When $\phi^{(0)}$ is inserted on the left or right in a 3-point function the correlator is unambiguous.
\item
When $\phi^{(0)}$ is inserted in the middle the correlator is ambiguous for $0 < \chi < 1$.  However it is only a binary ambiguity, since any particular path for analytic continuation
will give an overall $i \epsilon$ that agrees with either $i\epsilon_{12}$ or $i\epsilon_{13}$.
\end{itemize}
This makes it very tempting to interpret the ambiguity as a breakdown of operator associativity for $\phi^{(0)}$ in the region $0 < \chi < 1$.\footnote{Note that this region in the bulk
depends on the positions of both boundary operators.}  That is, when the bulk field is on the left or right we have unambiguous correlators
\be
\langle \phi^{(0)} {\cal O}_2 {\cal O}_3 \rangle \qquad {\rm and} \qquad \langle {\cal O}_2 {\cal O}_3 \phi^{(0)} \rangle
\ee
But if the bulk field is in the middle we need to specify an order of multiplication and the two possibilities are
\be
\langle \left({\cal O}_2 \phi^{(0)}\right) {\cal O}_3 \rangle \qquad {\rm and} \qquad \langle {\cal O}_2 \left(\phi^{(0)} {\cal O}_3\right) \rangle
\ee
The prescription for calculating $\langle \left({\cal O}_2 \phi^{(0)}\right) {\cal O}_3 \rangle$ is to analytically continue into $0 < \chi < 1$ starting from a point on the boundary
that is timelike to ${\cal O}_2$, and the prescription for calculating $\langle {\cal O}_2 \left(\phi^{(0)} {\cal O}_3\right) \rangle$ is to start from a point on the boundary that is
timelike to ${\cal O}_3$.

One way to argue for this interpretation is to compute the correlation function while multiplying the operators in a definite order.  We can do this by taking an OPE limit.  For example let's work
with spacelike-separated points on the boundary and work in the limit $\vert X_3 \vert \rightarrow \infty$.  Then we can compute the correlator by taking an OPE between $\phi^{(0)}$ and ${\cal O}_2$.
In this limit, as we show in appendix \ref{sect:GeneralDimensions}, we can continue into $0 < \chi < 1$ with a well-defined $i \epsilon$ prescription.  It is the same $i \epsilon$ that we would have
gotten by starting at timelike separation to ${\cal O}_2$.  We regard the OPE as multiplying ${\cal O}_2$ by $\phi^{(0)}$ first, then later multiplying by ${\cal O}_3$.  This supports the interpretation
as calculating $\langle \left({\cal O}_2 \phi^{(0)}\right) {\cal O}_3 \rangle$.  Of course we could have sent $\vert X_2 \vert \rightarrow \infty$ and started by taking the OPE between $\phi^{(0)}$ and
${\cal O}_3$.  We would interpret this as multiplying operators in the opposite order.\footnote{Our prescriptions for defining $\langle \left({\cal O}_2 \phi^{(0)}\right) {\cal O}_3 \rangle$ and
$\langle {\cal O}_2 \left(\phi^{(0)} {\cal O}_3\right) \rangle$ would also allow us to take OPE's at timelike separation.}

This has an amusing implication for commutators.  One might be tempted to interpret the prescription for defining $\langle \left({\cal O}_2 \phi^{(0)}\right) {\cal O}_3 \rangle$ as using a smearing
region for $\phi^{(0)}$ that can include ${\cal O}_2$ but is spacelike separated from ${\cal O}_3$, so that $\phi^{(0)}$ commutes with ${\cal O}_3$.  This is indeed what (\ref{IVcommutator}) suggests.
Likewise one might be tempted to interpret the prescription for defining $\langle {\cal O}_2 \left(\phi^{(0)} {\cal O}_3\right) \rangle$ as a smearing that can include ${\cal O}_3$ but is spacelike separated from ${\cal O}_2$, so that $\phi^{(0)}$ commutes with ${\cal O}_2$.  This is indeed what (\ref{Vcommutator}) suggests.  But if $\phi^{(0)}$ is taken to be non-associative then a direct attempt
to check this interpretation is foiled.  The parenthesis must go around the commutator, so for $0 < \chi < 1$ we have\footnote{Note that this would not have been possible for
a linear operator whose 3-point function with two other operators had only one branch cut.  With only one discontinuity across a cut available, one or the other of these commutators
would have to vanish.} 
\bea
\label{NonAssoc[]1}
&&\langle \left([{\cal O}_2, \phi^{(0)}]\right) {\cal O}_3 \rangle \not= 0 \\
\label{NonAssoc[]2}
&&\langle {\cal O}_2 \left([\phi^{(0)}, {\cal O}_3]\right) \rangle \not=0
\eea
We obtain the pleasing result that $\phi^{(0)}$, thought of as a non-associative operator, treats the other operators in a correlator democratically.

To conclude, in the language of quantum error correction we are asserting that $\phi^{(0)}$ is a logical but non-associative operator.  This means it cannot be interpreted in standard quantum mechanics.
The bulk reconstruction procedure corrects $\phi^{(0)}$ to become a well-defined CFT operator, which in our interpretation means that it restores associativity.  Somewhat surprisingly it also gives rise
to local bulk physics.  However the procedure for correcting $\phi^{(0)}$ (bulk reconstruction) only appears to be possible in $1/N$ perturbation theory, and cannot be carried out outside the code
subspace.  The reason for this is well-known.  The towers of higher-dimension multi-trace operators one needs to correct $\phi^{(0)}$ are independent operators in the $1/N$ expansion but not at
finite $N$.  So at finite $N$, or when the number of insertions in a correlator scales like some power of $N$, one cannot expect to be able to correct $\phi^{(0)}$ to get unambiguous bulk correlators.
In perturbation theory around a semiclassical background we can recover correlation functions from the CFT which look like they come from a local gauge-fixed bulk theory, but beyond
perturbation theory bulk observables are at best non-associative.  This means that at finite $N$ we cannot use bulk observables to build a bulk Hilbert space.  It appears that bulk quantum mechanics
only arises in perturbation theory around a semiclassical background.  It would be interesting to explore the relevance of this for the firewall paradox \cite{Almheiri:2012rt,Almheiri:2013hfa,Papadodimas:2012aq,Papadodimas:2013jku,Papadodimas:2015jra} and to understand if there is a connection to the obstacles to defining observables studied in \cite{Berenstein:2016pcx,Jafferis:2017tiu}.

\bigskip
\goodbreak
\centerline{\bf Acknowledgements}
\noindent
The work of DK is supported by U.S.\ National Science Foundation grant PHY-1519705.  The work of GL is supported in part by the Israel Science
Foundation under grants 504/13 and 447/17.

\appendix
\section{Construction of local bulk operators\label{sect:review}}
We give a quick review of the established procedure for constructing local bulk scalar fields in terms of the CFT.
We start with a primary scalar field ${\cal O}_\Delta$ of dimension $\Delta$ in the CFT and introduce a smearing function $K_{\Delta}(T,X,Z \vert T',X')$.
The smearing function should satisfy the free wave equation
\begin{equation}
\left(\nabla^{2}_{AdS}-\Delta(\Delta-d)\right) K_{\Delta}(T,X,Z \vert T',X') =0
\end{equation}
We define a free bulk field $\phi^{(0)}(T,X,Z)$ by
\begin{equation}
\phi^{(0)}(T,X,Z)=\int dT' d^{d-1}X' \, K_{\Delta}(T,X,Z \vert T',X') {\cal O}_{\Delta}(T',X')
\end{equation}
The smearing function should satisfy appropriate boundary conditions so that we recover ${\cal O}_\Delta$ from $\phi^{(0)}$ as the bulk point approaches the boundary.
The choice of smearing function is not unique, but a convenient choice is to take
\begin{equation}
\phi^{(0)}(T,X,Z) = {\Gamma(\Delta - {d \over 2}) \over 2 \pi^{d/2} \Gamma(\Delta-d+1)}\int\limits_{T'{}^2 + \vert Y' \vert^2 < Z^2} dT'd^{d-1}Y' \left(\frac{Z^2 - T'^2 - \vert Y' \vert^2}{Z}\right)^{\Delta-d} {\cal O}_{\Delta}(T+T',X+iY')
\end{equation}
where we have normalized
\be
\phi^{(0)}(T,X,Z) \sim {Z^\Delta \over 2 \Delta - d} {\cal O}(T,X) \qquad \hbox{\rm as $Z \rightarrow 0$}
\ee
This can also be written as an infinite sum.  For example in AdS${}_3$, with $T^\pm$ denoting light-front coordinates on the boundary, we have
\begin{equation}
\phi^{(0)}(T^+,T^-,Z)={Z^\Delta \over 2} \sum_{m=0}^{\infty}\frac{\Gamma(\Delta-1)Z^{2m}}{\Gamma(m+1)\Gamma(m + \Delta)}(\partial_+\partial_-)^{m}{\cal O}(T^+,T^-).
\label{sumop}
\end{equation}

The two-point function computed in the CFT, $\langle \phi^{(0)} \phi^{(0)} \rangle$, reproduces the bulk two-point function of a scalar field in AdS.  If one computes the 3-point function of $\phi^{(0)}$
with two boundary operators one gets 
\begin{equation}
\langle\phi^{(0)}(x,z) {\cal O}_{1}(y_1) {\cal O}_{2}(y_2)\rangle =\frac{1}{(y_1 - y_2)^{2\Delta_1}}
\left[\frac{z}{z^2+(x-y_2)^2}\right]^{\Delta_2-\Delta_1} I(\chi)
\label{3point11}
\end{equation}
Here $x$, $y_1$, $y_2$ are spacetime coordinates in the CFT.  We've introduced an AdS-invariant combination
\be
\label{chidef}
\chi=\frac{[(x-y_1)^2+z^2][(x-y_2)^2+z^2]}{(y_1-y_2)^2 z^2}
\ee
and
\begin{equation}
\label{Idef}
I(\chi) = \frac{\gamma}{2\Delta-d} \left(\frac{1}{\chi-1}\right)^\DS F\big(\,\DS,\,\DS - \frac{d}{2} + 1,\, \Delta - \frac{d}{2} + 1,\,\frac{1}{1-\chi}\,\big)
\end{equation}
where $\gamma$ is the coefficient of the CFT 3-point function and $\Delta_{\textstyle *}=\frac{1}{2}(\Delta+\Delta_{1}-\Delta_{2})$.

This result has singularities when the bulk point coincides with a bulk space-like geodesic connecting the boundary points $y_1$ and $y_2$, and it has a branch cut when the bulk point is to the future or past of this geodesic. Since this singularity appears when the bulk point is space-like separated from the boundary operators the correlation function does not obey bulk micro-causality. This was used
in \cite{Kabat:2015swa} as a guiding principle to reconstruct the bulk field. Micro-causality was restored by correcting the definition of the bulk field, adding an infinite sum of smeared double-trace
scalar primaries ${\cal O}^{1,2}_{\Delta_n}$ built from ${\cal O}_{1}$, ${\cal O}_{2}$ and $2n$ derivatives.
\begin{equation}
\phi(T,X,Z)=\phi^{(0)}+\frac{1}{N}\sum_{n} a_{n} \int dT' d^{d-1}X' \, K_{\Delta_{n}}(T,X,Z \vert T',X') {\cal O}^{1,2}_{\Delta_{n}}(T',X')
\end{equation}
The coefficients $a_{n}$ are to be chosen so that when $\phi$ is inserted inside a 3-point function in place of $\phi^{(0)}$, the 3-point function becomes analytic when the bulk point is space-like
separated from the boundary points. This leads to explicit expressions for $a_{n}$ which are unique up to field redefinitions.  The resulting bulk field $\phi$ has been shown to solve the expected
local bulk equation of motion to order $1/N$.

\section{Timelike separated boundary points\label{sect:timelike}}
In section \ref{sect:spacelike} we analyzed the 3-point correlator of a massless field in AdS${}_3$ with two boundary operators when the boundary operators were spacelike separated.
Here we study what happens when the boundary operators are timelike separated.  The upshot is that although the geometry is different, the qualitative outcome is the same.

\begin{figure}
\center{\includegraphics{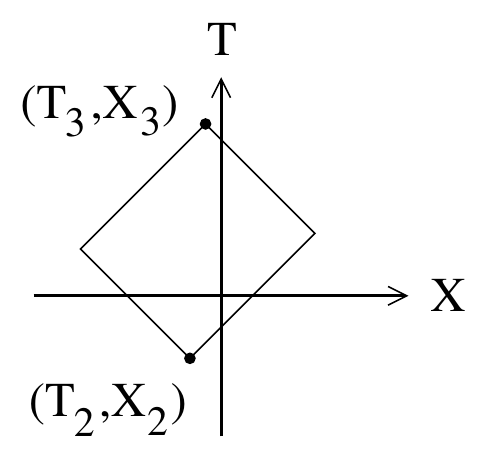} \hfil \includegraphics{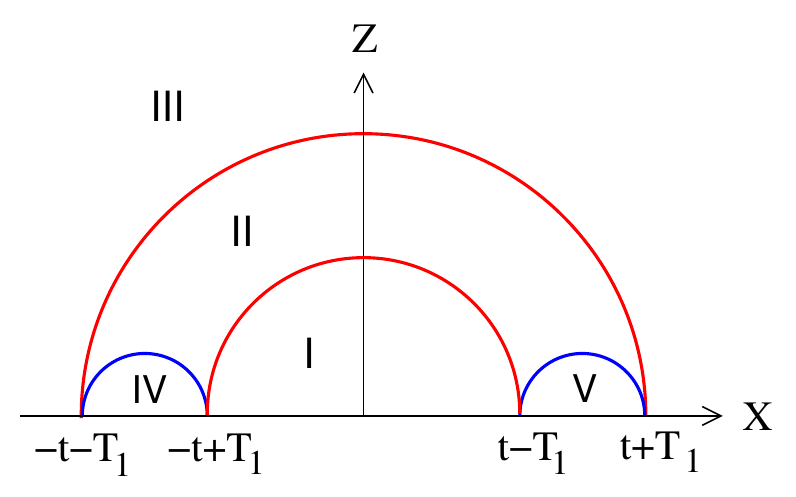}}
\caption{On the left, the causal diamond on the boundary determined by two timelike separated points.  On the right, a bulk time slice with various regions indicated.
The two red curves correspond to $\chi = 0$ and the two blue curves correspond to $\chi = 1$.\label{fig:timelike}}
\end{figure}

A pair of timelike separated points defines a causal diamond on the boundary as shown in the left panel
Fig.\ \ref{fig:timelike}.  From (\ref{ChiOverChi-1}) we see that now $\chi = 0$ corresponds to bulk points that are null separated from the top or bottom of the causal diamond,
while $\chi = 1$ corresponds to bulk points that are null separated from the left or right tip of the diamond.  So relative to the previous discussion, all that happens on a fixed time slice
of the bulk is that the red and blue curves get switched.  This is illustrated in the right panel of Fig.\ \ref{fig:timelike}, where without loss of generality we have taken the boundary
operators to be located at $(T_2 = -t,\,X_2 = 0)$ and $(T_3 = +t,\,X_3 = 0)$.  We take the bulk point to be $(T_1,X_1,Z_1)$, and for simplicity we restrict our analysis to $0 < T_1 < t$.

\begin{table}
\begin{center}
\begin{tabular}{c|c|l}
start from region & cross & \quad resulting $i \, {\rm Im} \, {\rm Arg}$ \\
\hline
\vphantom{\Big(}{\sf IV} & $\chi = 1$ at $C = 0$ & $i \epsilon_{12}(T_1 - t - X_1 + i \epsilon_{13}(T_1 + t + X_1)$ \\
{\sf V} & $\chi = 1$ at $D = 0$ & $i \epsilon_{12}(T_1 - t + X_1) + i \epsilon_{13}(T_1 + t - X_1)$ \\
{\sf I} & $\chi = 0$ at $B = 0$ & $-2 i \epsilon_{13}(T_1 - t)$ \\
{\sf III} & $\chi = 0$ at $A = 0$ & $2 i \epsilon_{12}(T_1 + t)$
\end{tabular}
\end{center}
\caption{Continuing into region {\sf II} for the case of timelike separated boundary operators.\label{TimelikeEpsilons}}
\end{table}

\newpage
With timelike separated boundary points, continuing into region {\sf II} gives the results shown in Table \ref{TimelikeEpsilons}.
Up to rescaling by position-dependent but positive factors this amounts to
\begin{center}
\begin{tabular}{c|c}
start from region & $i \, {\rm Im} \, {\rm Arg}$ \\
\hline
\vphantom{\Big(}{\sf IV} & $i \epsilon_{12} + i \epsilon_{13}$ \\
{\sf V} & $i \epsilon_{12} + i \epsilon_{13}$ \\
{\sf I} & $i \epsilon_{13}$ \\
{\sf III} & $i \epsilon_{12}$
\end{tabular}
\end{center}
The outcome is qualitatively the same as spacelike separation.  If the bulk operator is on the left or right in the correlator then the $\epsilon_{ij}$'s all have the same sign and the
continuation into region {\sf II} is unambiguous.  But if the bulk operator is in the middle, with say $\epsilon_{12} < 0$ and $\epsilon_{13} > 0$, then the result depends
on the starting region.  If we begin in region {\sf I} we get ${\rm Im} \, {\rm Arg} = 0^+$, while if we begin in region {\sf III}
we get ${\rm Im} \, {\rm Arg} = 0^-$.  The continuation from regions {\sf IV} and {\sf V} is ambiguous since it depends on the relative size of $\epsilon_{12}$ and $\epsilon_{13}$ as well as on the values of the coordinates $T_1,X_1,t$ at the point where the singularity is crossed.

\section{General spacetime and conformal dimensions\label{sect:GeneralDimensions}}
In this section we study the correlator for general conformal and spacetime dimensions.  The upshot is that the phenomena we found for massless fields in AdS${}_3$ in fact hold in general.

The 3-point function $\langle \phi^{(0)} {\cal O}_2 {\cal O}_3 \rangle$ for general operator dimensions $\Delta_1$, $\Delta_2$, $\Delta_3$ is given in (\ref{3point11}).  The result has singularities
at $\chi = 0$ and $\chi = 1$, where the AdS-invariant combination $\chi$ is in general defined by\footnote{In AdS${}_3$ one can write ${\chi \over \chi - 1}$ in the more instructive form (\ref{ChiOverChi-1}).}
\be
\label{GeneralChi}
\chi = {\left(-T_{12}^2 + \vert X_{12} \vert^2 + Z_1^2 \right) \, \left(-T_{13}^2 + \vert X_{13} \vert^2 + Z_1^2\right) \over \left(-T_{23}^2 + \vert X_{23} \vert^2\right) \, Z_1^2}
\ee
The first step is understanding where these singularities occur in the bulk.  Clearly the $\chi = 0$ singularities occur on light-cones emanating from the boundary points.  To understand
the $\chi = 1$ singularities we consider two cases.

\begin{figure}[h]
\center{\raisebox{5mm}{\includegraphics[height=5cm]{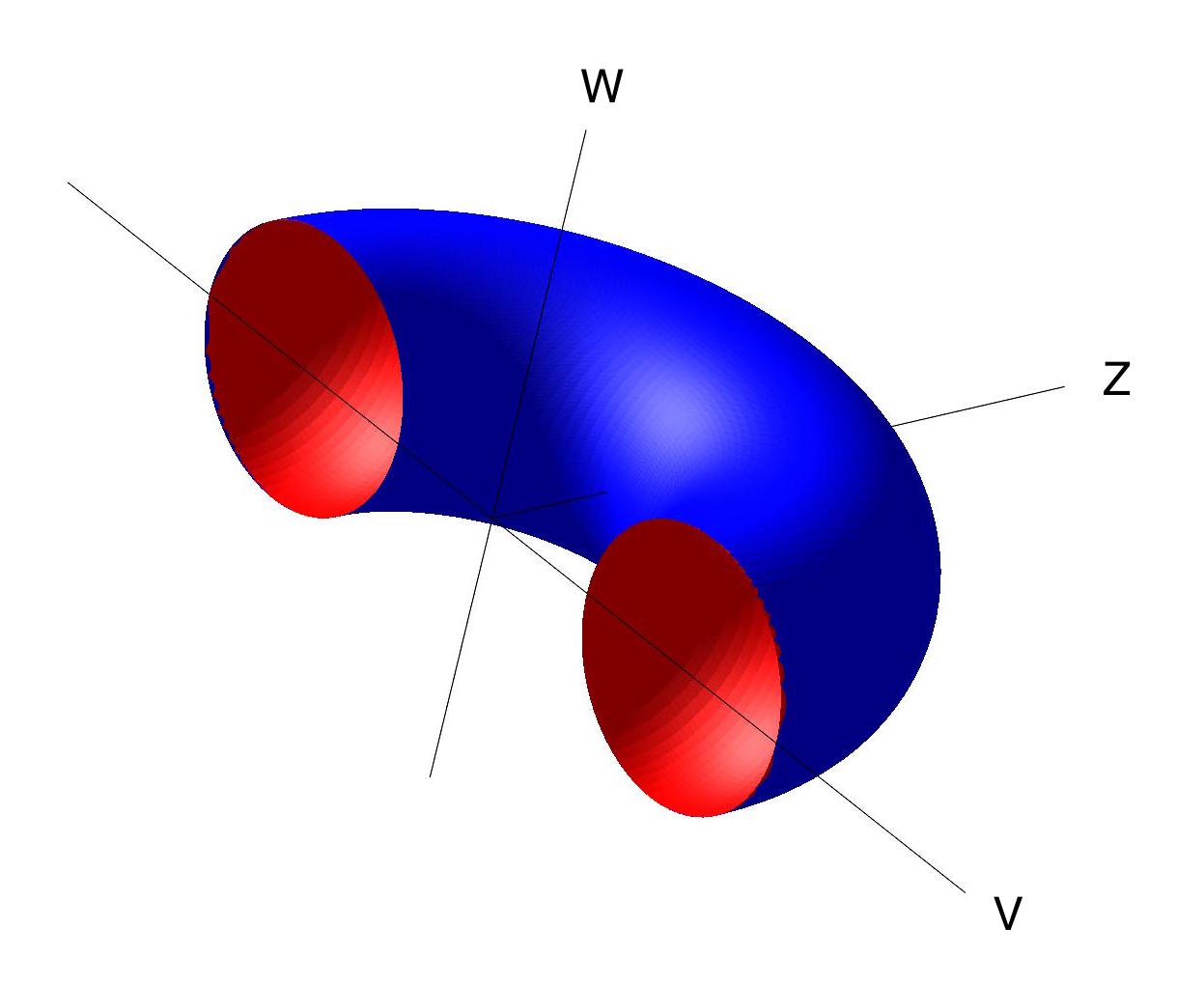}} \hspace{2.5cm} \includegraphics[height=6cm]{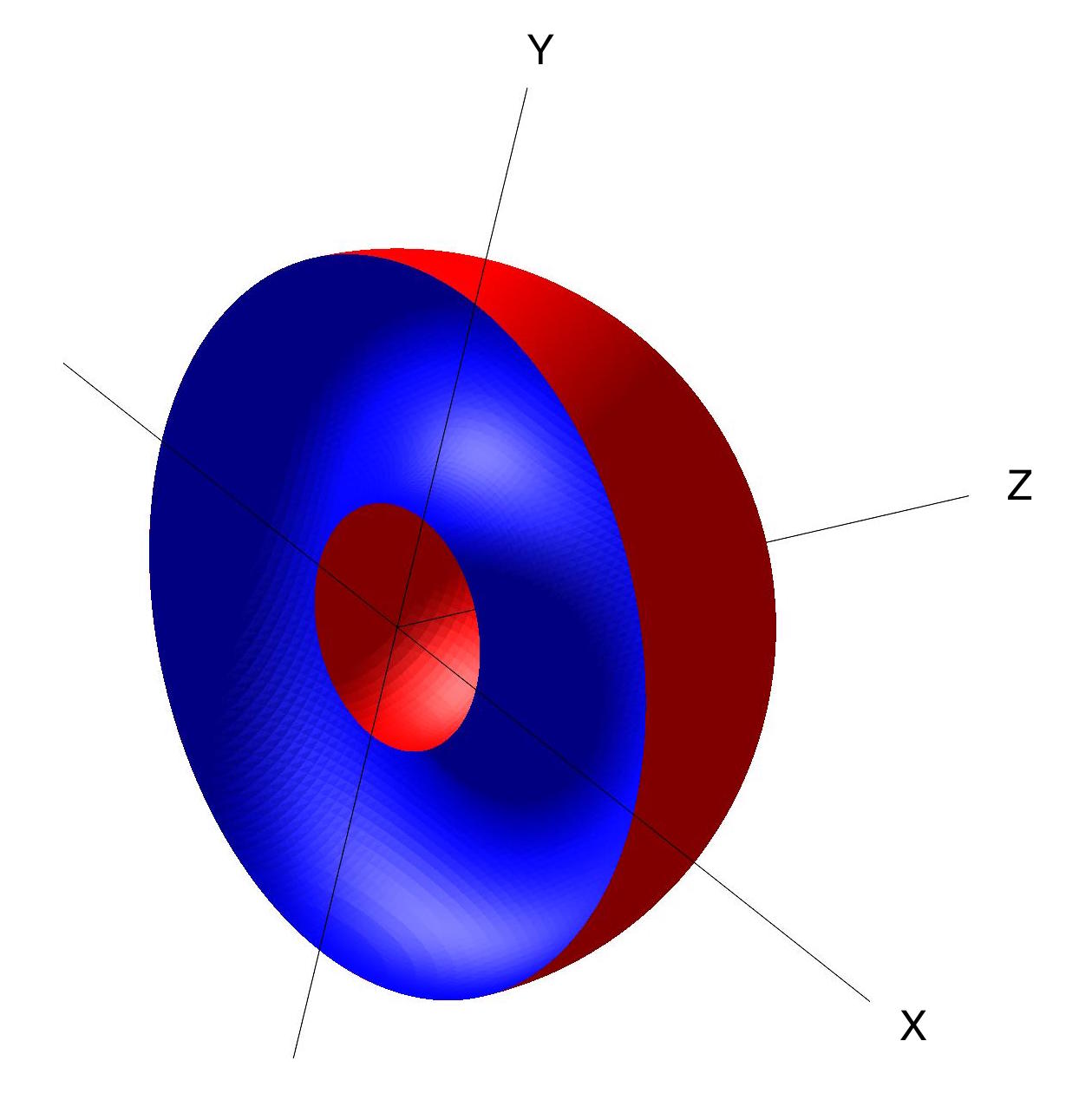}}
\caption{Left panel: a time slice for spacelike separated boundary points.  The blue tube is a shell of light expanding outward from the geodesic, corresponding to $\chi = 1$.
The red hemispheres are light shells emitted from the boundary points, corresponding to $\chi = 0$.  Right panel: a time slice for timelike separated boundary points.
The blue surface is a shell of light emitted from the waist of the boundary causal diamond, corresponding to $\chi = 1$.  The red hemispheres are light shells emitted from the past
(absorbed by the future) boundary points, corresponding to $\chi = 0$.\label{fig:ChiSurfaces}}
\end{figure}

\noindent
\underline{\em Spacelike separated boundary points} \\
\noindent
Without loss of generality we place the boundary points at equal times and we separate them along the first spatial coordinate of the CFT.  That is, we set
\be
T_2 = T_3 = 0 \qquad X_2 = (-y,\vec{0}\,) \quad X_3 = (+y,\vec{0}\,)
\ee
We label the bulk point $(T,X,Z)$ and look at the behavior on a bulk slice of fixed time with $0 < T <y$.  Decomposing the spatial coordinates of the bulk point into
their first component and the rest, $X = (V,\vec{W})$, the condition $\chi = 1$ has solutions
\be
V^2 + Z^2 = \left(y \pm \sqrt{T^2 - \vert \vec{W} \vert^2}\right)^2
\ee
Note that this requires $\vert \vec{W} \vert < T$.  At $T = 0$ we get a semicircle of radius $y$ in the $VZ$ plane.  This semicircle can be understood as a spacelike geodesic connecting the two boundary points.
For $T > 0$ the semicircle becomes a tube that expands transversely at the speed of light.  That is, the surface $\chi = 1$ is made up of light rays emitted perpendicular to the geodesic one has
at $T = 0$.  The tube ends on the boundary, where it meets the $\chi = 0$ surfaces as illustrated in the left panel of Fig.\ \ref{fig:ChiSurfaces}.

\noindent
\underline{\em Timelike separated boundary points} \\
\noindent
In this case we set
\be
T_2 = -t \quad T_3 = +t \qquad X_2 = X_3 = 0
\ee
and consider a bulk point with coordinates $(T,X,Z)$.
The condition $\chi = 1$ can be factored and has two branches of solutions.  On a bulk time slice with $-t < T < t$ only one branch is realized, and setting $\chi = 1$ implies that
\be
\left(\vert X \vert - t\right)^2 + Z^2 = T^2
\ee
When $T = 0$ we get a sphere of radius $t$ on the boundary.  This sphere can be thought of as the ``waist'' of the boundary causal diamond determined by the two boundary points,
meaning the intersection of the future lightcone of $(T_2,X_2)$ with the past lightcone of $(T_3,X_3)$.  For $T > 0$ the sphere expands transversely into the bulk at the speed of light.
This is illustrated in the right panel of Fig.\ \ref{fig:ChiSurfaces}.

Having understood where the singularities are located, now let's study the nature of the singularities.  To do this we introduce a Wightman $i\epsilon$ prescription and
replace $T_i \rightarrow T_i - i \epsilon_i$ in (\ref{GeneralChi}).  This gives $\chi$ an imaginary part, which resolves the singularity but in a way that depends on which surface we're crossing.  In general
\be
{\rm Im} \, \chi = {2 \epsilon_{12} T_{12}\left(-T_{13}^2 + X_{13}^2 + Z_1^2\right) + 2 \epsilon_{13} T_{13}\left(-T_{12}^2+X_{12}^2+Z_1^2\right) \over \left(-T_{23}^2 + X_{23}^2\right) \, Z_1^2}
\ee
Points with $\chi = 0$ are null separated from one or the other of the boundary points.  So when $\chi = 0$ one of the terms in the numerator vanishes and we have a well-defined
$i \epsilon$ prescription.  But at $\chi = 1$ the imaginary part depends on both $\epsilon_{12}$ and $\epsilon_{13}$ and the prescription for crossing the singularity is ambiguous.

There are two situations worth commenting on, in which the ambiguity associated with crossing a $\chi = 1$ singularity can be avoided.
\begin{itemize}
\item
Suppose one crosses $\chi = 1$, but in the limit where one approaches null separation from one of the boundary points.  Then the $i \epsilon$ associated with that boundary
point is the one that dominates.
\item
One can also work in an OPE limit where one sends one of the boundary points to infinity, in either a timelike or spacelike direction.  Then the $i \epsilon$ associated with the
remaining boundary point is the one that matters.
\end{itemize}

\begin{figure}
\center{\includegraphics{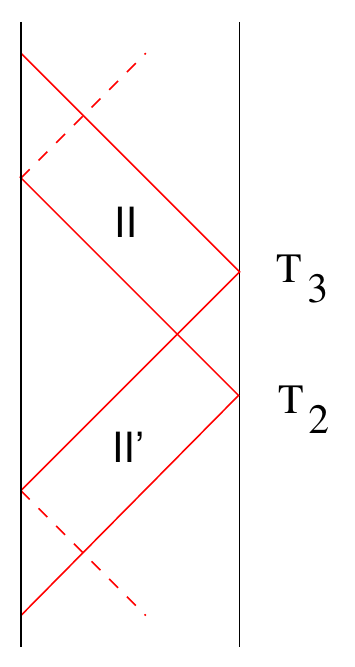}}
\caption{Penrose diagram for AdS${}_2$.  In AdS${}_2$ the only singularities occur when the bulk point is lightlike to one of the boundary points.
These light rays bounce off the boundary as indicated by dotted lines.  This makes correlators of $\phi^{(0)}$ in regions {\sf II} and {\sf II'}
ambiguous.\label{fig:AdS2}}
\end{figure}

Thus the outcome in arbitrary spacetime dimensions is qualitatively similar to what we found in AdS${}_3$.  There are a few distinct features that arise in other dimensions.
\begin{itemize}
\item
In AdS${}_4$ and higher, for spacelike separated boundary points, the region analogous to {\sf I} and {\sf III} in Fig.\ \ref{fig:spacelike} is connected.  Likewise for timelike separated boundary
points the region analogous to {\sf IV} and {\sf V} in Fig.\ \ref{fig:timelike} is connected.
\item
AdS${}_2$ is somewhat analogous to timelike separated boundary points in AdS${}_3$.  In AdS${}_2$ singularities only arise at null separation, so there are no $\chi = 1$ singularities to cross.   But there is still an obstacle to regarding $\phi^{(0)}$ as a well-defined bulk observable.  This can be seen from Fig.\ \ref{fig:AdS2}.  The ambiguity arises because to continue a correlator into region {\sf II} or {\sf II'} one must decide whether to
start from the left or right boundary.  These two different starting points yield $i \epsilon$ prescriptions which depend on which light cone is being crossed.  The different $i \epsilon$ prescriptions don't affect correlators
with $\phi^{(0)}$ on the left or right but give different results when $\phi^{(0)}$ is placed in the middle.
\end{itemize}


\begin{thebibliography}{10}

\bibitem{Maldacena:1997re}
J.~M. Maldacena, ``The large $N$ limit of superconformal field theories and
  supergravity,'' {\em Adv. Theor. Math. Phys.} {\bfseries 2} (1998) 231--252,
\href{http://arxiv.org/abs/hep-th/9711200}{{\ttfamily hep-th/9711200}}.

\bibitem{Kabat:2011rz}
D.~Kabat, G.~Lifschytz, and D.~A. Lowe, ``{Constructing local bulk observables
  in interacting AdS/CFT},''
  \href{http://dx.doi.org/10.1103/PhysRevD.83.106009}{{\em Phys.Rev.}
  {\bfseries D83} (2011) 106009},
\href{http://arxiv.org/abs/1102.2910}{{\ttfamily arXiv:1102.2910 [hep-th]}}.

\bibitem{Banks:1998dd}
T.~Banks, M.~R. Douglas, G.~T. Horowitz, and E.~J. Martinec, ``{AdS dynamics
  from conformal field theory},''
\href{http://arxiv.org/abs/hep-th/9808016}{{\ttfamily arXiv:hep-th/9808016}}.

\bibitem{Dobrev:1998md}
V.~K. Dobrev, ``{Intertwining operator realization of the AdS/CFT
  correspondence},''
  \href{http://dx.doi.org/10.1016/S0550-3213(99)00284-9}{{\em Nucl. Phys.}
  {\bfseries B553} (1999) 559--582},
\href{http://arxiv.org/abs/hep-th/9812194}{{\ttfamily arXiv:hep-th/9812194}}.

\bibitem{Bena:1999jv}
I.~Bena, ``{On the construction of local fields in the bulk of AdS(5) and other
  spaces},'' \href{http://dx.doi.org/10.1103/PhysRevD.62.066007}{{\em Phys.
  Rev.} {\bfseries D62} (2000) 066007},
\href{http://arxiv.org/abs/hep-th/9905186}{{\ttfamily arXiv:hep-th/9905186}}.

\bibitem{Hamilton:2005ju}
A.~Hamilton, D.~Kabat, G.~Lifschytz, and D.~A. Lowe, ``Local bulk operators in
  AdS/CFT: A boundary view of horizons and locality,'' {\em Phys. Rev.}
  {\bfseries D73} (2006) 086003,
\href{http://arxiv.org/abs/hep-th/0506118}{{\ttfamily hep-th/0506118}}.

\bibitem{Hamilton:2006az}
A.~Hamilton, D.~Kabat, G.~Lifschytz, and D.~A. Lowe, ``Holographic
  representation of local bulk operators,'' {\em Phys. Rev.} {\bfseries D74}
  (2006) 066009,
\href{http://arxiv.org/abs/hep-th/0606141}{{\ttfamily hep-th/0606141}}.

\bibitem{Hamilton:2006fh}
A.~Hamilton, D.~Kabat, G.~Lifschytz, and D.~A. Lowe, ``Local bulk operators in
  AdS/CFT: A holographic description of the black hole interior,'' {\em Phys.
  Rev.} {\bfseries D75} (2007) 106001,
\href{http://arxiv.org/abs/hep-th/0612053}{{\ttfamily hep-th/0612053}}.

\bibitem{Verlinde:2015qfa}
H.~Verlinde, ``{Poking holes in AdS/CFT: Bulk fields from boundary states},''
\href{http://arxiv.org/abs/1505.05069}{{\ttfamily arXiv:1505.05069 [hep-th]}}.

\bibitem{Miyaji:2015fia}
M.~Miyaji, T.~Numasawa, N.~Shiba, T.~Takayanagi, and K.~Watanabe, ``{Continuous
  multiscale entanglement renormalization ansatz as holographic surface-state
  correspondence},''
  \href{http://dx.doi.org/10.1103/PhysRevLett.115.171602}{{\em Phys. Rev.
  Lett.} {\bfseries 115} no.~17, (2015) 171602},
\href{http://arxiv.org/abs/1506.01353}{{\ttfamily arXiv:1506.01353 [hep-th]}}.

\bibitem{Nakayama:2015mva}
Y.~Nakayama and H.~Ooguri, ``{Bulk locality and boundary creating operators},''
  \href{http://dx.doi.org/10.1007/JHEP10(2015)114}{{\em JHEP} {\bfseries 10}
  (2015) 114},
\href{http://arxiv.org/abs/1507.04130}{{\ttfamily arXiv:1507.04130 [hep-th]}}.

\bibitem{Kabat:2017mun}
D.~Kabat and G.~Lifschytz, ``{Local bulk physics from intersecting modular
  Hamiltonians},'' \href{http://dx.doi.org/10.1007/JHEP06(2017)120}{{\em JHEP}
  {\bfseries 06} (2017) 120},
\href{http://arxiv.org/abs/1703.06523}{{\ttfamily arXiv:1703.06523 [hep-th]}}.

\bibitem{Faulkner:2017vdd}
T.~Faulkner and A.~Lewkowycz, ``{Bulk locality from modular flow},''
  \href{http://dx.doi.org/10.1007/JHEP07(2017)151}{{\em JHEP} {\bfseries 07}
  (2017) 151},
\href{http://arxiv.org/abs/1704.05464}{{\ttfamily arXiv:1704.05464 [hep-th]}}.

\bibitem{Cotler:2017erl}
J.~Cotler, P.~Hayden, G.~Salton, B.~Swingle, and M.~Walter, ``{Entanglement
  wedge reconstruction via universal recovery channels},''
\href{http://arxiv.org/abs/1704.05839}{{\ttfamily arXiv:1704.05839 [hep-th]}}.

\bibitem{Kabat:2015swa}
D.~Kabat and G.~Lifschytz, ``{Bulk equations of motion from CFT correlators},''
  \href{http://dx.doi.org/10.1007/JHEP09(2015)059}{{\em JHEP} {\bfseries 09}
  (2015) 059},
\href{http://arxiv.org/abs/1505.03755}{{\ttfamily arXiv:1505.03755 [hep-th]}}.

\bibitem{Heemskerk:2012mn}
I.~Heemskerk, D.~Marolf, J.~Polchinski, and J.~Sully, ``{Bulk and transhorizon
  measurements in AdS/CFT},''
  \href{http://dx.doi.org/10.1007/JHEP10(2012)165}{{\em JHEP} {\bfseries 1210}
  (2012) 165},
\href{http://arxiv.org/abs/1201.3664}{{\ttfamily arXiv:1201.3664 [hep-th]}}.

\bibitem{Lewkowycz:2016ukf}
A.~Lewkowycz, G.~J. Turiaci, and H.~Verlinde, ``{A CFT perspective on
  gravitational dressing and bulk locality},''
  \href{http://dx.doi.org/10.1007/JHEP01(2017)004}{{\em JHEP} {\bfseries 01}
  (2017) 004},
\href{http://arxiv.org/abs/1608.08977}{{\ttfamily arXiv:1608.08977 [hep-th]}}.

\bibitem{Guica:2016pid}
M.~Guica, ``{Bulk fields from the boundary OPE},''
\href{http://arxiv.org/abs/1610.08952}{{\ttfamily arXiv:1610.08952 [hep-th]}}.

\bibitem{Anand:2017dav}
N.~Anand, H.~Chen, A.~L. Fitzpatrick, J.~Kaplan, and D.~Li, ``{An exact
  operator that knows its location},''
\href{http://arxiv.org/abs/1708.04246}{{\ttfamily arXiv:1708.04246 [hep-th]}}.

\bibitem{Chen:2017dnl}
H.~Chen, A.~L. Fitzpatrick, J.~Kaplan, and D.~Li, ``{The AdS$_3$ propagator and
  the fate of locality},''
\href{http://arxiv.org/abs/1712.02351}{{\ttfamily arXiv:1712.02351 [hep-th]}}.

\bibitem{Kabat:2012av}
D.~Kabat and G.~Lifschytz, ``{CFT representation of interacting bulk gauge
  fields in AdS},'' \href{http://dx.doi.org/10.1103/PhysRevD.87.086004}{{\em
  Phys.Rev.} {\bfseries D87} (2013) 086004},
\href{http://arxiv.org/abs/1212.3788}{{\ttfamily arXiv:1212.3788 [hep-th]}}.

\bibitem{Kabat:2013wga}
D.~Kabat and G.~Lifschytz, ``{Decoding the hologram: Scalar fields interacting
  with gravity},'' \href{http://dx.doi.org/10.1103/PhysRevD.89.066010}{{\em
  Phys.Rev.} {\bfseries D89} no.~6, (2014) 066010},
\href{http://arxiv.org/abs/1311.3020}{{\ttfamily arXiv:1311.3020 [hep-th]}}.

\bibitem{Heemskerk:2012np}
I.~Heemskerk, ``{Construction of bulk fields with gauge redundancy},''
  \href{http://dx.doi.org/10.1007/JHEP09(2012)106}{{\em JHEP} {\bfseries 1209}
  (2012) 106},
\href{http://arxiv.org/abs/1201.3666}{{\ttfamily arXiv:1201.3666 [hep-th]}}.

\bibitem{Kabat:2012hp}
D.~Kabat, G.~Lifschytz, S.~Roy, and D.~Sarkar, ``{Holographic representation of
  bulk fields with spin in AdS/CFT},''
  \href{http://dx.doi.org/10.1103/PhysRevD.86.026004,
  10.1103/PhysRevD.86.029901}{{\em Phys.Rev.} {\bfseries D86} (2012) 026004},
\href{http://arxiv.org/abs/1204.0126}{{\ttfamily arXiv:1204.0126 [hep-th]}}.

\bibitem{Sarkar:2014dma}
D.~Sarkar and X.~Xiao, ``{Holographic representation of higher spin gauge
  fields},'' \href{http://dx.doi.org/10.1103/PhysRevD.91.086004}{{\em
  Phys.Rev.} {\bfseries D91} no.~8, (2015) 086004},
\href{http://arxiv.org/abs/1411.4657}{{\ttfamily arXiv:1411.4657 [hep-th]}}.

\bibitem{Almheiri:2014lwa}
A.~Almheiri, X.~Dong, and D.~Harlow, ``{Bulk locality and quantum error
  correction in AdS/CFT},''
  \href{http://dx.doi.org/10.1007/JHEP04(2015)163}{{\em JHEP} {\bfseries 04}
  (2015) 163},
\href{http://arxiv.org/abs/1411.7041}{{\ttfamily arXiv:1411.7041 [hep-th]}}.

\bibitem{Mintun:2015qda}
E.~Mintun, J.~Polchinski, and V.~Rosenhaus, ``{Bulk-boundary duality, gauge
  invariance, and quantum error corrections},''
  \href{http://dx.doi.org/10.1103/PhysRevLett.115.151601}{{\em Phys. Rev.
  Lett.} {\bfseries 115} no.~15, (2015) 151601},
\href{http://arxiv.org/abs/1501.06577}{{\ttfamily arXiv:1501.06577 [hep-th]}}.

\bibitem{Ryu:2006bv}
S.~Ryu and T.~Takayanagi, ``{Holographic derivation of entanglement entropy
  from AdS/CFT},'' \href{http://dx.doi.org/10.1103/PhysRevLett.96.181602}{{\em
  Phys. Rev. Lett.} {\bfseries 96} (2006) 181602},
\href{http://arxiv.org/abs/hep-th/0603001}{{\ttfamily arXiv:hep-th/0603001
  [hep-th]}}.

\bibitem{Hijano:2015zsa}
E.~Hijano, P.~Kraus, E.~Perlmutter, and R.~Snively, ``{Witten diagrams
  revisited: The AdS geometry of conformal blocks},''
  \href{http://dx.doi.org/10.1007/JHEP01(2016)146}{{\em JHEP} {\bfseries 01}
  (2016) 146},
\href{http://arxiv.org/abs/1508.00501}{{\ttfamily arXiv:1508.00501 [hep-th]}}.

\bibitem{Czech:2016xec}
B.~Czech, L.~Lamprou, S.~McCandlish, B.~Mosk, and J.~Sully, ``{A stereoscopic
  look into the bulk},'' \href{http://dx.doi.org/10.1007/JHEP07(2016)129}{{\em
  JHEP} {\bfseries 07} (2016) 129},
\href{http://arxiv.org/abs/1604.03110}{{\ttfamily arXiv:1604.03110 [hep-th]}}.

\bibitem{daCunha:2016crm}
B.~Carneiro~da Cunha and M.~Guica, ``{Exploring the BTZ bulk with boundary
  conformal blocks},''
\href{http://arxiv.org/abs/1604.07383}{{\ttfamily arXiv:1604.07383 [hep-th]}}.

\bibitem{deBoer:2016pqk}
J.~de~Boer, F.~M. Haehl, M.~P. Heller, and R.~C. Myers, ``{Entanglement,
  holography and causal diamonds},''
  \href{http://dx.doi.org/10.1007/JHEP08(2016)162}{{\em JHEP} {\bfseries 08}
  (2016) 162},
\href{http://arxiv.org/abs/1606.03307}{{\ttfamily arXiv:1606.03307 [hep-th]}}.

\bibitem{Kabat:2016zzr}
D.~Kabat and G.~Lifschytz, ``{Locality, bulk equations of motion and the
  conformal bootstrap},'' \href{http://dx.doi.org/10.1007/JHEP10(2016)091}{{\em
  JHEP} {\bfseries 10} (2016) 091},
\href{http://arxiv.org/abs/1603.06800}{{\ttfamily arXiv:1603.06800 [hep-th]}}.

\bibitem{Almheiri:2012rt}
A.~Almheiri, D.~Marolf, J.~Polchinski, and J.~Sully, ``{Black holes:
  Complementarity or firewalls?},''
  \href{http://dx.doi.org/10.1007/JHEP02(2013)062}{{\em JHEP} {\bfseries 1302}
  (2013) 062},
\href{http://arxiv.org/abs/1207.3123}{{\ttfamily arXiv:1207.3123 [hep-th]}}.

\bibitem{Almheiri:2013hfa}
A.~Almheiri, D.~Marolf, J.~Polchinski, D.~Stanford, and J.~Sully, ``{An
  apologia for firewalls},''
  \href{http://dx.doi.org/10.1007/JHEP09(2013)018}{{\em JHEP} {\bfseries 09}
  (2013) 018},
\href{http://arxiv.org/abs/1304.6483}{{\ttfamily arXiv:1304.6483 [hep-th]}}.

\bibitem{Papadodimas:2012aq}
K.~Papadodimas and S.~Raju, ``{An infalling observer in AdS/CFT},''
  \href{http://dx.doi.org/10.1007/JHEP10(2013)212}{{\em JHEP} {\bfseries 1310}
  (2013) 212},
\href{http://arxiv.org/abs/1211.6767}{{\ttfamily arXiv:1211.6767 [hep-th]}}.

\bibitem{Papadodimas:2013jku}
K.~Papadodimas and S.~Raju, ``{State-dependent bulk-boundary maps and black
  hole complementarity},''
  \href{http://dx.doi.org/10.1103/PhysRevD.89.086010}{{\em Phys.Rev.}
  {\bfseries D89} no.~8, (2014) 086010},
\href{http://arxiv.org/abs/1310.6335}{{\ttfamily arXiv:1310.6335 [hep-th]}}.

\bibitem{Papadodimas:2015jra}
K.~Papadodimas and S.~Raju, ``{Comments on the necessity and implications of
  state-dependence in the black hole interior},''
\href{http://arxiv.org/abs/1503.08825}{{\ttfamily arXiv:1503.08825 [hep-th]}}.

\bibitem{Berenstein:2016pcx}
D.~Berenstein and A.~Miller, ``{Can topology and geometry be measured by an
  operator measurement in quantum gravity?},''
  \href{http://dx.doi.org/10.1103/PhysRevLett.118.261601}{{\em Phys. Rev.
  Lett.} {\bfseries 118} no.~26, (2017) 261601},
\href{http://arxiv.org/abs/1605.06166}{{\ttfamily arXiv:1605.06166 [hep-th]}}.

\bibitem{Jafferis:2017tiu}
D.~L. Jafferis, ``{Bulk reconstruction and the Hartle-Hawking wavefunction},''
\href{http://arxiv.org/abs/1703.01519}{{\ttfamily arXiv:1703.01519 [hep-th]}}.

\end{thebibliography}
\providecommand{\href}[2]{#2}\begingroup\raggedright\endgroup

\end{document}